\documentclass[useAMS,usenatbib]{mn2e}
\usepackage{amssymb, amsmath}
\usepackage{graphicx}
\usepackage{subfigure}
\usepackage{multirow}
\usepackage{hyperref}
\usepackage{color}

\title[X-ray Reverberation in Mrk 335]{Simultaneous spectral and reverberation modelling of relativistic reflection in Mrk 335}

\author[P. Chainakun and A. J. Young]{P. Chainakun and A. J. Young \\
H. H. Wills Physics Laboratory, Tyndall Avenue, Bristol BS8 1TL}

\begin{document}

\date{}

\pagerange{\pageref{firstpage}--\pageref{lastpage}} \pubyear{2014}

\maketitle

\label{firstpage}

\begin{abstract}
We present an X-ray spectral and timing model to investigate the broad and variable iron line seen in the high flux state of Mrk 335. The model consists of a variable X-ray source positioned along the rotation axis of the black hole that illuminates the accretion disc producing a back-scattered, ionized reflection spectrum. We compute time lags including full dilution effects and perform simultaneous fitting of the 2--10 keV spectrum and the frequency-dependent time lags of 2.5--4 vs. 4--6.5 keV bands. The best-fitting parameters are consistent with a black hole mass of $\approx 1.3 \times 10^{7} M_{\odot}$, disc inclination of $45^{\circ}$ and the photon index of the direct continuum of 2.4. The iron abundance is 0.5 and the ionization parameter is $10^{3}$ erg cm s$^{-1}$ at the innermost part of the disc and decreases further out. The X-ray source height is very small, $\approx 2 r_{g}$. Furthermore, we fit the Fe L lags simultaneously with the 0.3--10 keV spectrum. The key parameters are comparable to those previously obtained. We also report the differences below 2 keV using the {\sc xillver} and {\sc reflionx} models which could affect the interpretation of the soft excess. While simultaneously fitting spectroscopic and timing data can break the degeneracy between the source height and the black hole mass, we find that the measurements of the source height and the central mass significantly depend on the ionization state of the disc and are possibly model-dependent.

\end{abstract}


\section{Introduction}

The narrow line Seyfert 1 galaxy Mrk 335 ($z = 0.0258$) has been observed by several X-ray observatories over a range of spectral states. In 2006, \emph{XMM-Newton} found that in a high flux state Mrk 335 exhibits very clear double-peaked Fe K$\alpha$ and Fe XXVI Ly$\alpha$ emission lines around 6.4 and 7 keV, respectively \citep{On07, La08}. The best-fitting model revealed evidence of relativistically blurred inner disc reflection but the narrow peaks at 6.4 and 7 keV are due to the reflection from more distant material such as the molecular torus, and optically thin ionized gas filling the torus, respectively. In 2007, the source went into an extremely low flux state in which the overall flux dropped by a factor of $\approx 10$ and the spectrum was successfully described by either partially covering absorption or blurred reflection models \citep{Gr07, Gr08}. In 2009 when the source was in an intermediate flux state, the spectral fitting required only the blurred reflection from the disc around a rapidly spinning black hole without invoking partial covering \citep{Gr12, Ga13}. Recently, \cite{Pa14} studied the spectrum and emissivity profile of Mrk 335 by using the \emph{NuSTAR} data from 2013 observations and constrain the spin parameter of the central super-massive black hole to be $a \approx 0.98$.

Alongside time-averaged spectral data analysis, timing properties of Mrk 335 have been studied. Similar to most AGN, it has positive time lags at low frequencies in which higher energy bands lag lower energy bands. \cite{Ar08} found that the sharp cut-off in the lags occurred at a frequency close to the bend frequency in the power spectrum which could be described by a model of inward propagating fluctuations \citep[e.g.,][]{Ly97, Ko01, Ar06}. This model assumes the X-ray emission hardens towards the centre, so lower frequency variability in soft emission at larger radii propagates inwards and modulates the higher frequency variability in hard emission at smaller radii, thereby producing the positive hard lags. Additionally, many AGN with rapid variability show negative soft lags at high frequencies, referred to as reverberation lags (i.e. the delay between changes in the X-ray continuum and the associated echoes from the disc). Since the reflected photons travel an additional distance to the disc before being observed, the soft reflection variations lag behind the primary hard-band variations, with the amplitude of the lag being associated with the light travel time between the X-ray source and the disc \citep[see][for a review]{Ut14}. Measuring reverberation lags provides a new method to probe the location/size of the X-ray source. The first hint of a negative lag was seen by \citet{McHardy2007}, with the first robust detection of the Fe L lags being in 1H0707-495 \citep{Fa09}. After that the Fe L lags have been detected in many AGN including Mrk 335 \citep[e.g.,][]{De13}. Furthermore, \cite{Ka13} found that the Fe K lags are also present in Mrk 335 during the high flux state. These lags disappeared once the source went into an intermediate flux state. Combining the high and medium flux state data showed lags are comparable to those from the high flux state alone, demonstrating that the lag signal is dominated by the high flux data.

The interplay between the disc reflection and propagating fluctuations seems to provide a complete explanation of time lags in AGN. This framework is preferred over alternative models \citep[e.g., reflection from  distant materials proposed by][]{Mi10} because it can explain, for example, the absence of energy dependent Fe K lags at low frequencies \citep{Ka13} as the low- and high-frequency lags are generated by two distinct mechanisms. The reverberation occurs at shorter timescales compared to the inwards propagating fluctuations so they dominate at higher frequencies while the propagation lags take over at lower frequencies.

In this paper we analyse the high flux data from \emph{XMM-Newton} during observations in 2006. We focus on self-consistently modelling time-averaged \emph{and} and lag-frequency spectra of Mrk 335 in the lamp-post geometry scenario (i.e., the disc is illuminated by an axial isotropic point source). We first investigate the lags between the hard bands, 2.5--4 vs. 4--6.5 keV, before shifting the lags to two softer bands, 0.3--0.8 vs. 1--4 keV. The photon trajectories are traced along null geodesics in the Kerr metric between the source, the disc and the observer \citep{Ba72, Cu75, Ka92, Re99, Ru00, Ch12}. We assume the disc to be geometrically thin, optically thick \citep{Sh73} and apply the {\sc xillver} model \citep{Gar10, Gar13} to deal with the X-ray reprocessing by the disc via fluorescence, Compton-scattering and photoelectric absorption. The ionization state of the disc is determined given the disc density and the illuminating flux. The total spectrum is divided into two components: Power-Law dominated Component (PLC) and Reflection Dominated Component (RDC). In principle the RDC lags behind the PLC since it is dominated by reflected photons which take a longer time before being observed. The reflection responses to the primary variations (so-called response functions) are used to compute their delays in the Fourier frequency domain, indicative of reverberation lags. We parametrise the positive continuum lags by a power-law, as in \cite{Em14}.

Nevertheless, each energy band contains both power-law and reflection components. The measured lags then are diluted as there is some continuum emission in the RDC and some reflection emission in the PLC \citep{Wi13}. Recently \cite{Ca14} investigated the effects of key parameters such as the source height, inclination angle and the black hole mass on time lags. They applied the reflected response fraction to the RDC to dilute the Fe K$\alpha$ lags of NGC 4151, and assume no contribution of reflection to the PLC. \cite{Em14} successfully fitted the Fe L lags of 12 AGN by computational modelling the reverberation lags and employing a power-law for the positive lags. They, however, assumed the contamination of the continuum in the RDC and no reflection flux in the PLC. Here we include the full contamination between cross-components (soft in the hard band and hard in the soft band) so that full dilution effects are taken into account and, as a result, the lags are further suppressed. As our model predicts time-averaged energy spectra and frequency-dependent reverberation lags, the low frequency positive lags are modelled separately by a phenomenologically motivated power-law. All fitting was performed using {\sc ISIS} \citep{Ho00}.

The remainder of this paper is organised as follows. We describe the \emph{XMM-Newton} observations and data reduction in Section~\ref{sec:data_reduction}. The reverberation model based on ray tracing simulation is presented in Section~\ref{sec:ray_tracing}. The model is investigated for cases in which there is a density gradient in the ionized disc. In Section~\ref{sec:model_fits}, we present the model parameters for Mrk 335 and perform simultaneous fitting of the time-averaged and lag-frequency spectra. The conclusions are presented in Section~\ref{sec:conclusions}. We discuss the differences we find when using the {\sc reflionx} model \citep{Ge91, Ro99, Ro05} in the Appendix.

\section{Observations} \label{sec:data_reduction}

Mrk 335 was observed by \emph{XMM-Newton} \citep{Jansen2001} on 2006-01-03 for 133~ksec. The time lag and time-averaged spectra were produced in a standard way, following the data reduction of \cite{Ka13} using the \emph{XMM-Newton} Science Analysis System version 14.0.0. The data were screened for background flares, and filtered using the conditions {\tt PATTERN <= 4} and {\tt FLAG == 0}, leaving 120~ksec of good data. The source spectrum and light curves were extracted from a circular aperture of radius $35^{\prime\prime}$, and background counts were extracted from a similarly sized but offset aperture. Background subtracted lightcurves were produced using the {\tt lccorr} tool with 10~s time bins. The time lags were computed from the phase of the average cross power spectrum in the standard way, as were the error bars on time lags, following \citet{No99}. The time lags that we obtain are consistent with those previously published in the literature.

\section{Spectral and Timing Model} \label{sec:ray_tracing}
This section summarises essential equations and theoretical framework relating to our physical reverberation model. We will first describe the ray tracing simulation in Section 3.1 followed by the response function which is used to produce the full-contaminated light curve in Section 3.2. In Section 3.3 we investigate the reverberation lags and discuss the effects of dilution, ionization, disc density, and disc outer radius on the time lags and energy spectrum.
\subsection{Ray tracing and X-ray reprocessing}

The Kerr metric describing the spacetime around a rotating black hole is characterised by the black hole mass, $M$, and the angular momentum, $J$, parametrised in the form of the specific angular momentum or spin parameter $a=J/M$. Throughout the ray tracing simulation we use Boyer-Lindquist coordinates and measure the distance and time in gravitational units, $r_g = GM/c^2$ and $t_g = GM/c^3$, respectively. The disc is assumed to be a standard geometrically thin, optically thick disc \citep{Sh73} extending from the innermost stable circular orbit, $r_{ms}$, to $400r_g$. We assume the observer is at $1000 r_g$ from the black hole and consider a lamp-post geometry in which an isotropic point source is positioned at height $h$ along the symmetry axis.

To compute the direct and reflection spectra, an adaptive step-size method is used to trace all photon trajectories along the null geodesics between the source, the disc and the observer. The source spectrum is a powerlaw with flux density given by $F_s(E_s) \propto E_s^{-\alpha}$, where $E_s$ is the photon energy and $\alpha$ is the energy index. All conserved quantities are the total energy, $E$, the angular momentum, $L$, and the Carter constant, $Q$ \citep{Ca68}. The equations of motion of a photon are \citep{Ba72, Cu75, Ka92, Fa97, Re99}
\begin{align}
\Sigma \frac{dr}{d\lambda} & = \pm(V_r)^{1/2}, \\
\Sigma \frac{d\theta}{d\lambda} & = \pm(V_\theta)^{1/2}, \\
\Sigma \frac{d\varphi}{d\lambda} & = -(aE-L/\sin^2\theta)+aT/\Delta, \\
\Sigma \frac{dt}{d\lambda} & = -a(aE\sin^2\theta - L)+(r^2+a^2)T/\Delta,
\end{align}
where
\begin{align}
\Sigma & = r^2+a^2cos^{2}\theta, \\
\Delta & = r^2+a^2-2Mr, \\
T & = E(r^2+a^2)-La, \\
V_r & = T^2 - \Delta {[}{\mu^2}r^2 + (L-aE)^2+Q{]}, \\
V_{\theta} & = Q - \cos^2\theta {[}a^2(\mu^2-E^2)+L^2/\sin^2\theta {]},
\end{align}
Note that $\lambda=\tau/\mu$ where $\tau$ and $\mu$ are the proper time and the rest mass of the test particle, respectively.

We perform photon path integrals numerically in parallel on NVIDIA K20 Graphic Processing Unit (GPU) cards on the BlueCrystal Supercomputer at the University of Bristol. Those photons from the source that arrive at the observer directly are given the direct spectrum,
\begin{equation}
F_{D}(E_{o})=g_{so}^{\alpha+1}E_{o}^{-\alpha}d \Xi_{so} / dS_{o},
\end{equation}
where $d \Xi_{so}$ is the solid angle corresponding to the surface area $dS_{o}$ on the observer sky and $g_{so}$ is redshift of the source in the observer's frame. The redshift, $g_{so}$ is defined as
\begin{equation}
g_{so} = \frac{\nu_o}{\nu_s} = \bf{\frac{p_o  \cdot u_o}{p_s \cdot u_s}},
\end{equation}
where $\bf p_{o(s)}$ is the 4-momentum of the observed (source) photon and $\bf u_{o(s)}$ is the 4-velocity of the observer (source).

Furthermore, we divide the disc into small $drd\phi$-bins. The incident flux per unit area of each bin is then given by
\begin{equation}
F_{d}(E_{d})=g_{sd}^{\alpha+1}E_{d}^{-\alpha}d \Xi_{sd} / dS_{d},
\end{equation}
where $g_{sd}$ is redshift between the source and the disc, $dS_{d}$ is the unit surface area in the disc frame and $\Xi_{sd}$ is the solid angle corresponding to the disc element $drd\phi$ \citep[see][for a full expression]{Ru00, Dovciak2004, Ch12}. The corresponding ionization parameter is defined as
\begin{equation}
\xi(r,\varphi) = \frac{4\pi F_{t}(r,\varphi)}{n(r)},
\end{equation}
where $F_{t}(r,\varphi)$ is the total illuminating flux in that bin. The disc density is assumed to follow a power-law form: $n(r) \propto r^{-p}$. We apply the {\sc xillver} model \citep{Gar10, Gar13} to obtain the reflection spectrum allowing the iron abundance, $A_{Fe}$, and the photon index, $\Gamma$, to be model parameters. A comparison with the {\sc reflionx} model \citep{Ge91, Ro99, Ro05} will be given in the Appendix. Note that the photon index relates to the energy index via $\Gamma = \alpha + 1$.

To calculate the illumination of the disc we trace $\sim 5 \times 10^7$ photons from an isotropic point source to the disc as described by \citet{Ch12}. To trace the photons from the disc to the observer, we integrate photon paths backwards through $8192 \times 8192$ pixels on the observer's sky to the disc \citep{Fa97}. The reflection spectrum is blurred once it is transferred to the observer's frame. The total time-averaged spectrum is the sum of the direct and blurred reflection components.

\subsection{Response function and full, contaminated light curve}

\begin{figure}
\centering
\includegraphics*[width=70mm]{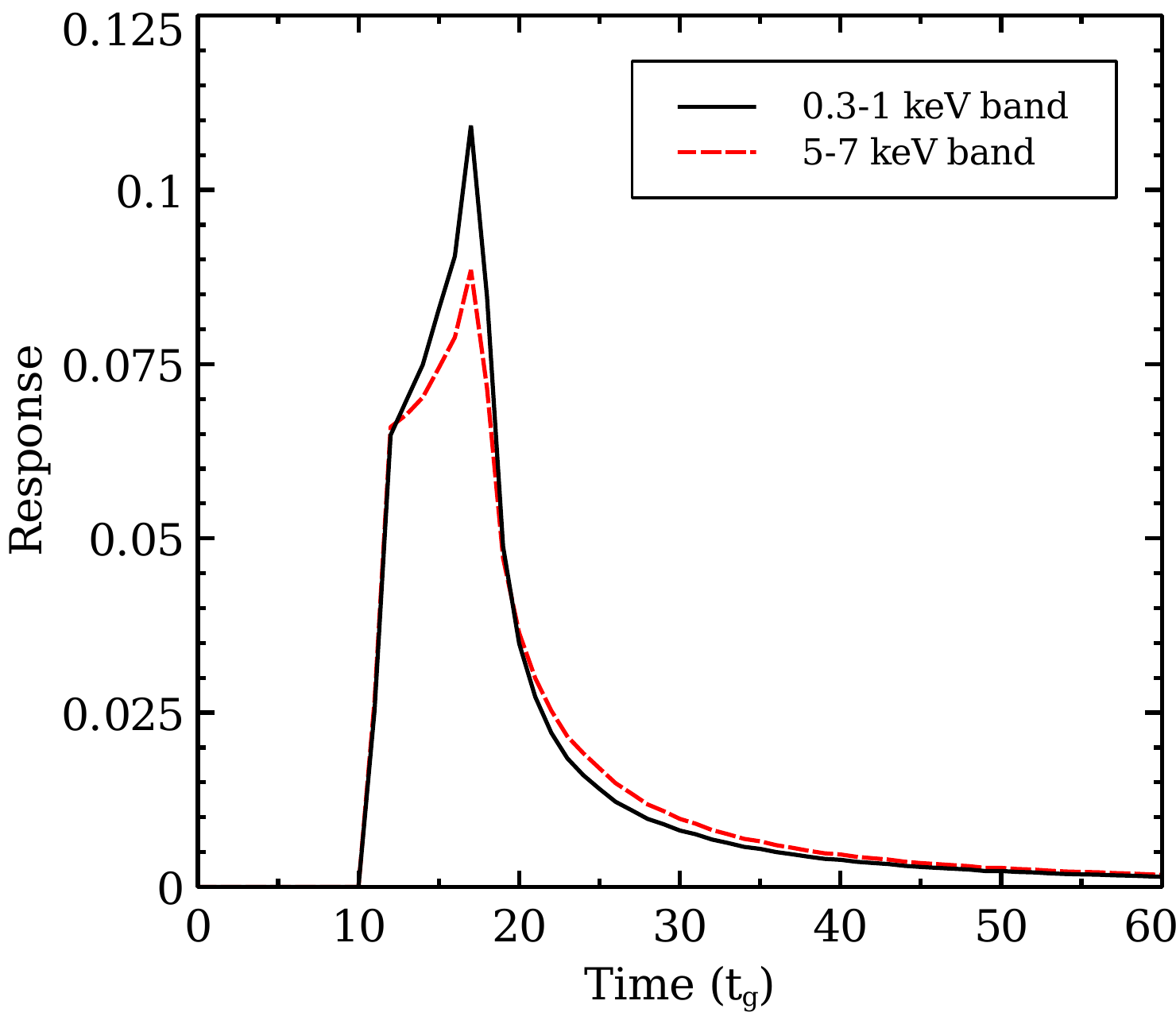}
\vspace{-0.3cm}
\caption{Response functions of the 0.3--1 keV (black, solid line) and 5--7 keV (red, dashed line) bands normalised so that their underlying areas equal 1. The relative changes of spectra between both bands for a disc with ionization gradient are not only in amplitude but also in shape which could lead to differences in the profiles. See text for more details. \label{fig:response_functions}}
\end{figure}

The response function, $\psi(t)$, which is used to produce time lags is the count rates of reflection photons as a function of time after the primary X-ray variations. However the reflection spectra from the disc could be different, resulting in different response profiles for the soft and hard energy bands, $\psi_S(t)$ and $\psi_H(t)$, respectively. Examples are shown in Fig.~\ref{fig:response_functions} for the responses of the 0.3--1 keV and 5--7 keV bands to an instantaneous flash of a point source at $h = 5 r_g$. Other parameters are iron abundance $A_{Fe}=1$, photon index $\Gamma=2$, spin $a=0.998$, inclination $i=30^\circ$ and ionization state at the innermost stable circular orbit $\xi_{ms} = 10^{3} \text{ erg cm s}^{-1}$. For a disc with ionization gradient the relative changes in spectra between each energy band are different corresponding to various ionization states on the disc. The differences in the responses are the result of the different energy bands selected. This implies the choice of the energy bands can affect the response function and hence the time lags. Throughout this paper we normalise the areas of the response functions to 1 and apply soft and hard reflected response fractions, $R_S$ and $R_H$, respectively, to determine the relative levels of their responses. These fractions also determine the levels of contamination in both bands which cause dilution effects and, as a result, the measured lags is less than the intrinsic lags.

Assuming $a(t)$ is the primary X-ray continuum flux as a function of time, we derive the full, contaminated soft and hard band light curves, $S(t)$ and $H(t)$, respectively, as
\begin{equation}
S(t) = a_{S}(t) + R_{S}\int_0^t a(t^\prime) \psi_{S}(t-t^\prime)dt^\prime
\end{equation}
and
\begin{equation}
H(t) = a_{H}(t) + R_{H}\int_0^t a(t^\prime) \psi_{H}(t-t^\prime)dt^\prime.
\end{equation}
The source flux can be varied to produce several ionization states on the disc. However, for simplicity, we select only one ionization profile each time when we do Fourier transform of the response function. In other words we assume the ionization gradient does not change over a period of time $t$ for the primary variation $a(t)$. Since the direct continuum follows the primary variations, $a_{S}(t)$ and $a_{H}(t)$ are interpreted as the contribution from the direct power-law components in soft and hard bands, respectively. The integral terms represent the response of the disc to earlier variations that contribute to the reflected flux. $R_S$ and $R_H$ are, respectively, the soft and hard reflected response fractions defined, in the same way as other authors, as $(\text{reflection flux}) / (\text{continuum flux})$ indicative the contamination ratio between cross-components. The fraction $R_S$ is one of our model parameters while $R_H$ is then measured from the corresponding spectra. Equations (14) and (15) are normalised such that $R_{S(H)}=1$ corresponds to equal contribution of the direct and reflection components in the soft (hard) band.

\subsection{Frequency-dependent time lags}

The nature of reverberation lags in AGN has been investigated through a wide range of parameters \citep{Wi13, Ca14, Em14}. These lags are sensitive to the source's height as they relate to the light-crossing timescales between the source and the disc. If the black hole mass increases, the lags are systematically shifted to lower frequencies and higher amplitudes \citep{De13}. The spin parameter and the inclination angle, however, seem to have less effect on the lags. See \cite{Ca14} for a detailed comparison of lags for different geometries of the accretion flows. Throughout this section we select the soft and hard bands to be 0.3--0.8 keV (RDC) and 1--4 keV (PLC) bands, respectively. When the power-law flux contaminates the RDC, the reflected response fraction in the RDC decreases and time lags decrease. On the other hand, when the reflection flux contaminates the PLC, the reflected response fraction in the PLC increases and time lags decrease. The more contamination between cross-components the more the time lags are diluted. Current literature such as \cite{Ca14} and \cite{Em14} include the direct flux in the RDC but assume no reflected flux in the PLC. Here we produce the full, diluted time lags by also including the reflection component in the PLC.

We investigate further the case of a density, and hence ionization, gradient in the disc and the dilution effects of contamination between the hard and soft bands. We follow \cite{No99} to calculate the frequency-dependent time lags by taking the Fourier transform of the full, contaminated light curves, $S(t)$ and $H(t)$. Their cross-spectrum is
\begin{equation}
C(f) = S^{*}(f)H(f).
\end{equation}
The argument of the cross-spectrum gives the phase difference, $\phi(f) = \arg C(f)$, which relates to time lags via
\begin{equation}
\tau(f) = \frac{\phi(f)}{2\pi f}.
\end{equation}
In our definition, a negative lag means the soft band lags behind the hard band. The phase difference is defined over $-\pi$ to $\pi$, so phase wrapping occurs at $f \ge 1 / 2\tau$ where the profile fluctuates around 0.

\begin{figure}
\centering
\includegraphics*[width=65mm]{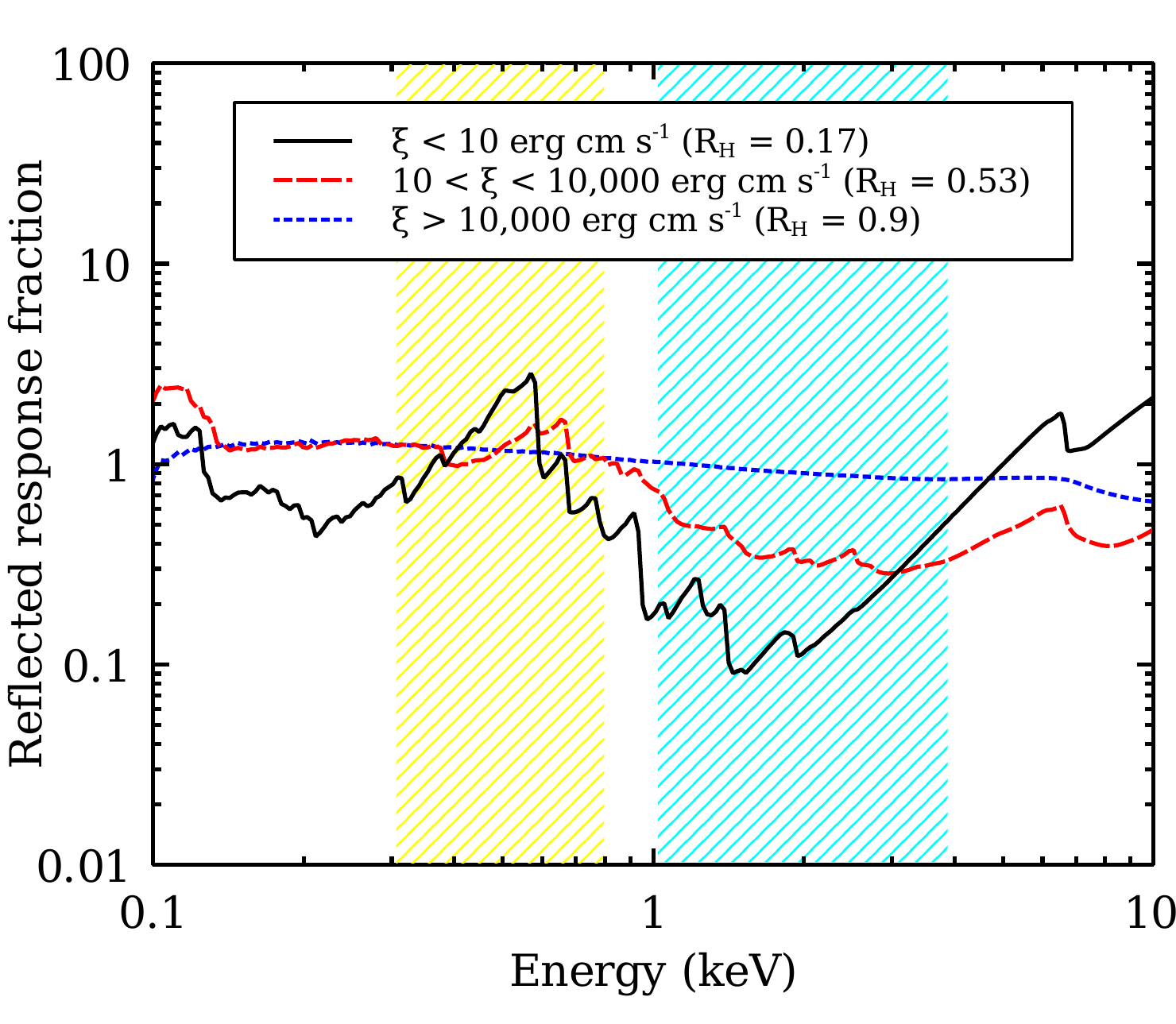}
\vspace{-0.3cm}
\includegraphics*[width=65mm]{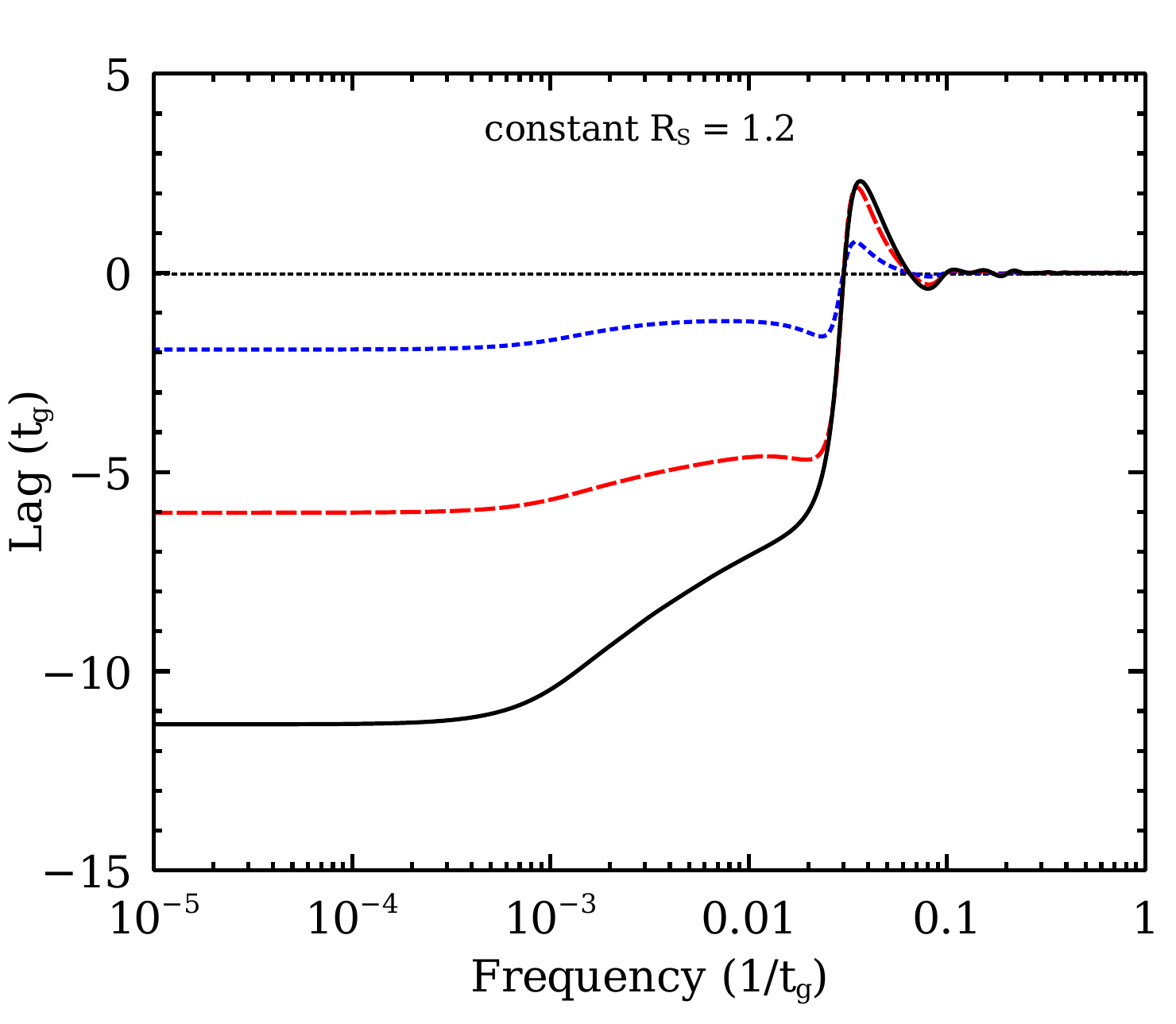}
\caption{\emph{Top panel:} The reflected response fraction for the cold (black line), medium-ionized (red, dashed line) and highly-ionized (blue, dotted line) disc. We allow the ionization states, given a value at the inner radius, to vary across the disc. The measured $R_H$ for fixing the $R_S = 1.2$ are 0.17, 0.53 and 0.9 respectively. The soft and hard bands are highlighted by yellow and blue background shading, respectively. \emph{Bottom panel:} The corresponding time lags as a function of Fourier frequency. As the disc becomes more highly ionized the amplitude of the time lag decreases. \label{fig:ref_fraction}}
\end{figure}

The necessity of including the full components in the soft and hard band light curves is emphasised by \cite{Wi13}. This is achievable by our model. The variations in flux-contamination can be physically produced when, for example, the ionization state of the disc changes in response to changes in the source luminosity. Fig.~\ref{fig:ref_fraction} represents the reflected response fraction and lag spectra for the source at $5r_{g}$ whose luminosity is varied to produce low, medium, and highly ionized discs. Fixing $R_{S}=1.2$, the measured $R_{H}$ is 0.17, 0.53 and 0.9, respectively. Other parameters are $i=30^{\circ}$, $A_{Fe}=1$ and $\Gamma=2$.  Even though the light-crossing distances are constant, the time lags decrease with increasing ionization state, and for the highly ionized disc the lags are very small. This is because when the disc becomes more ionized, the reflection spectra become flatter so $R_H$ increases and, as a result, the lag amplitude decreases.

\begin{figure}
\centering
\includegraphics*[width=65mm]{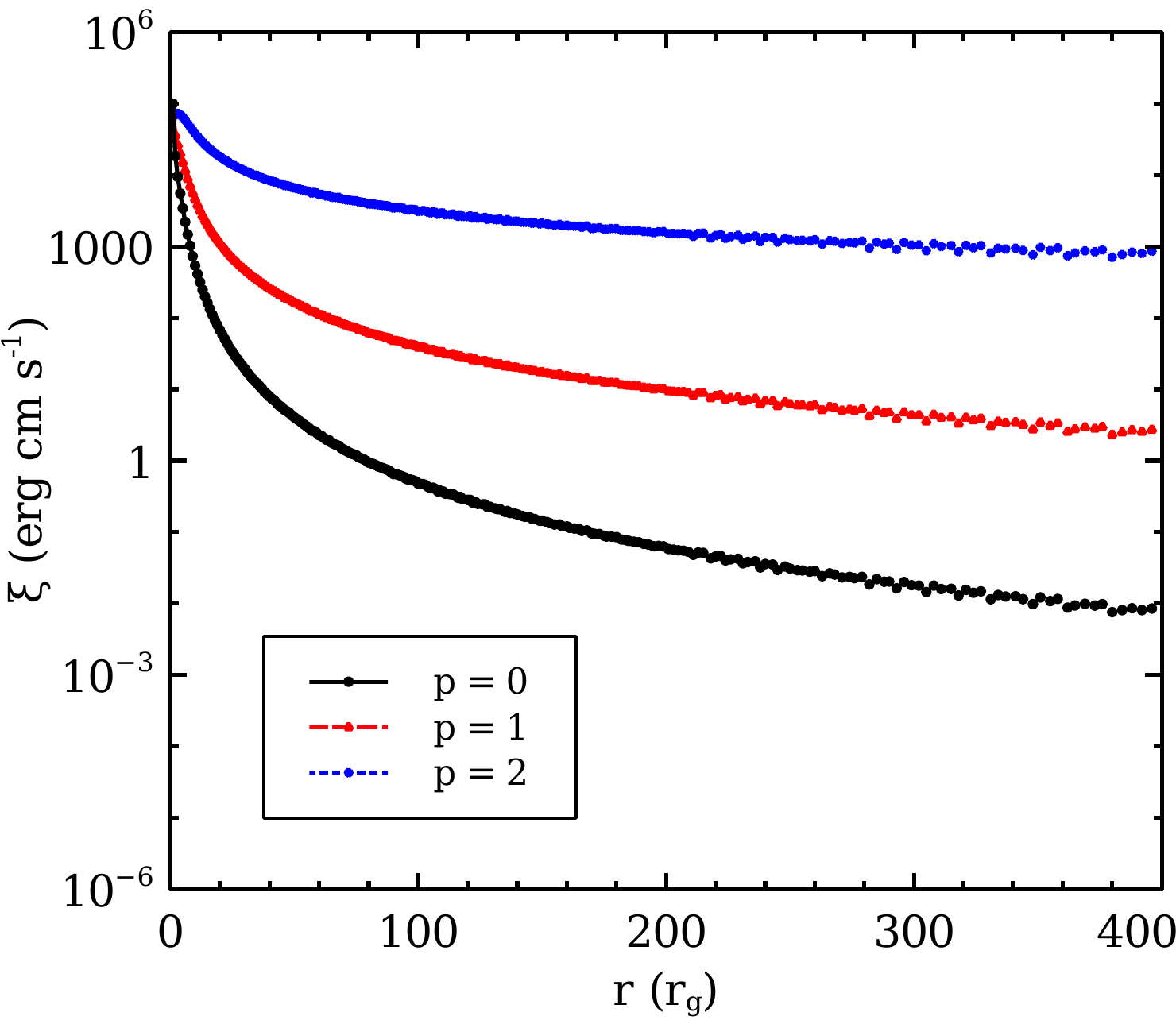}
\vspace{-0.1cm}
\includegraphics*[width=65mm]{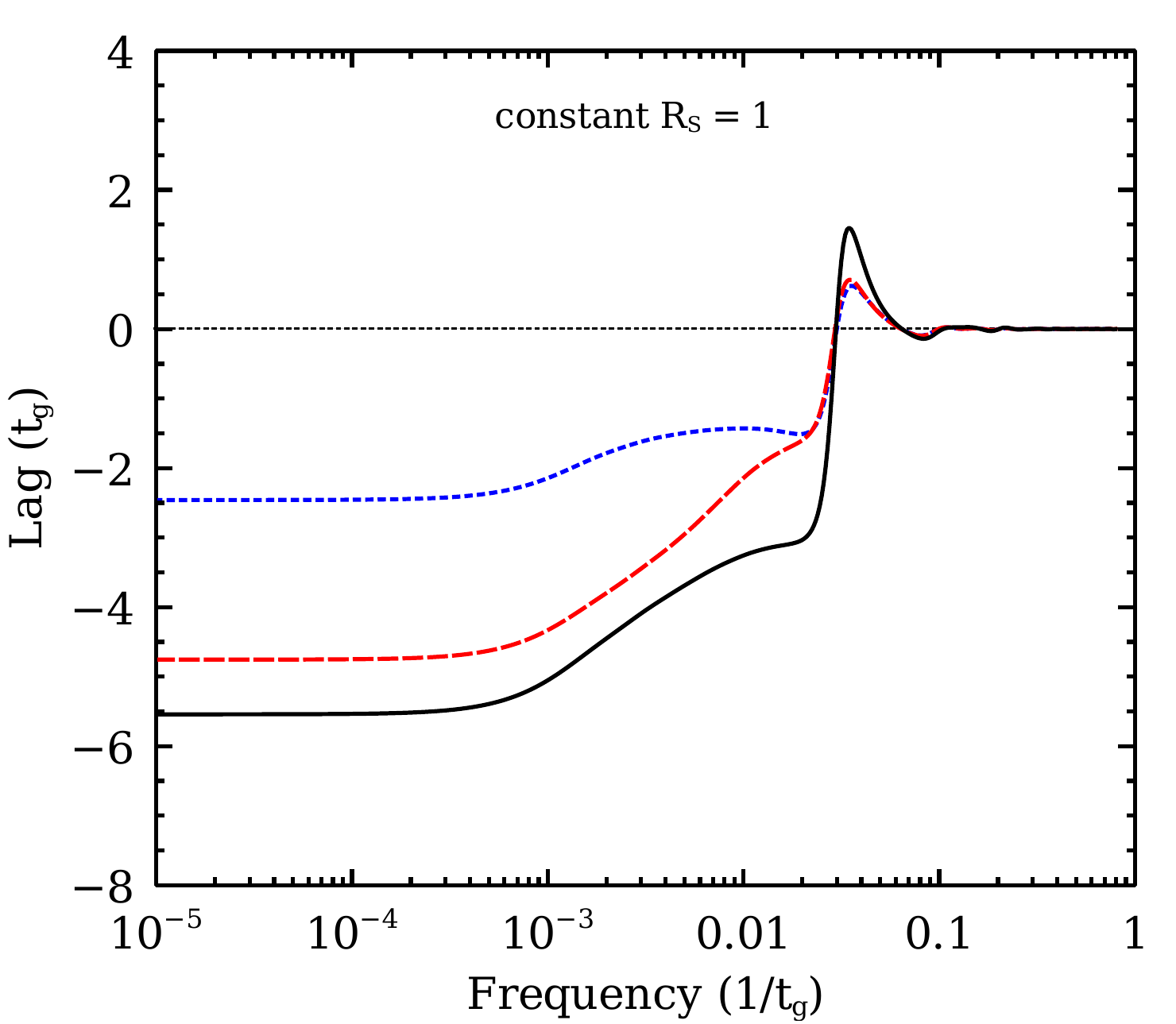}
\caption{\emph{Top panel:} ionization state of the disc for different density profiles, $n(r) \propto r^{-p}$. \emph{Bottom panel:} The corresponding time lags. In the scenario simulated here, discs with stronger density gradients, i.e., larger $p$, show shorter lags because the disc can remain highly ionized out to larger radii. \label{fig:density_index}}
\end{figure}

According to the model, the ionization state depends not only on the total incident flux, but also on the disc density which we assume to vary radially as $n(r) \propto r^{-p}$. Fig.~\ref{fig:density_index} shows how the ionization state and lags depend on the density index, $p$. For comparison, we fix the ionization state at the innermost radius, $\xi_{ms}$, to be $10^{5} \text{ erg cm s}^{-1}$. As the inner disc is more strongly illuminated than the outer parts, the ionization state should decrease further out. Decreasing $p$ increases the slope of ionization profile covering broader ranges down to very low states at the outer radii. This leads to the smaller $R_{H}$ and larger reverberation lags.

\begin{figure}
\centering
\includegraphics*[width=65mm]{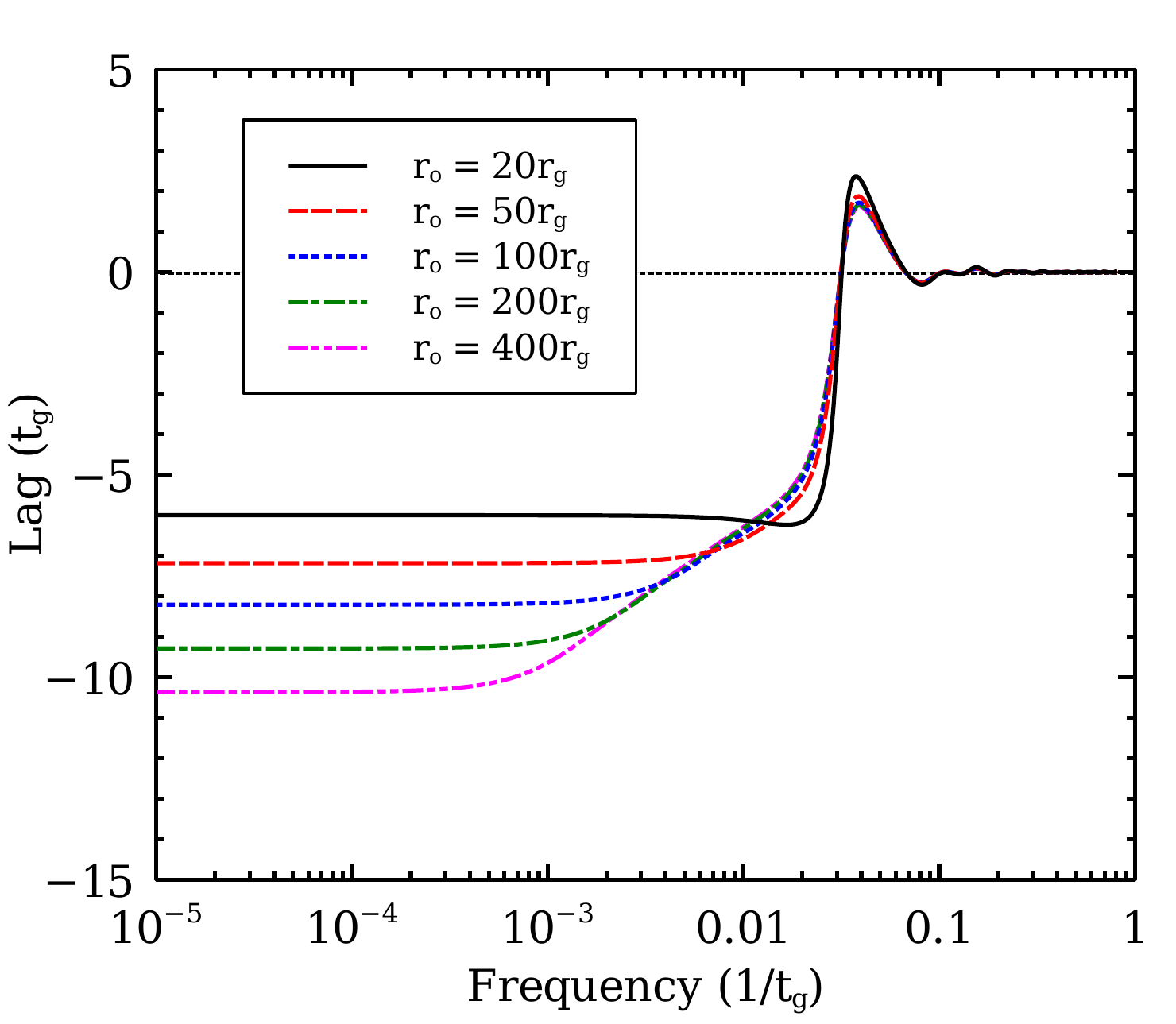}
\vspace{-0.1cm}
\includegraphics*[width=65mm]{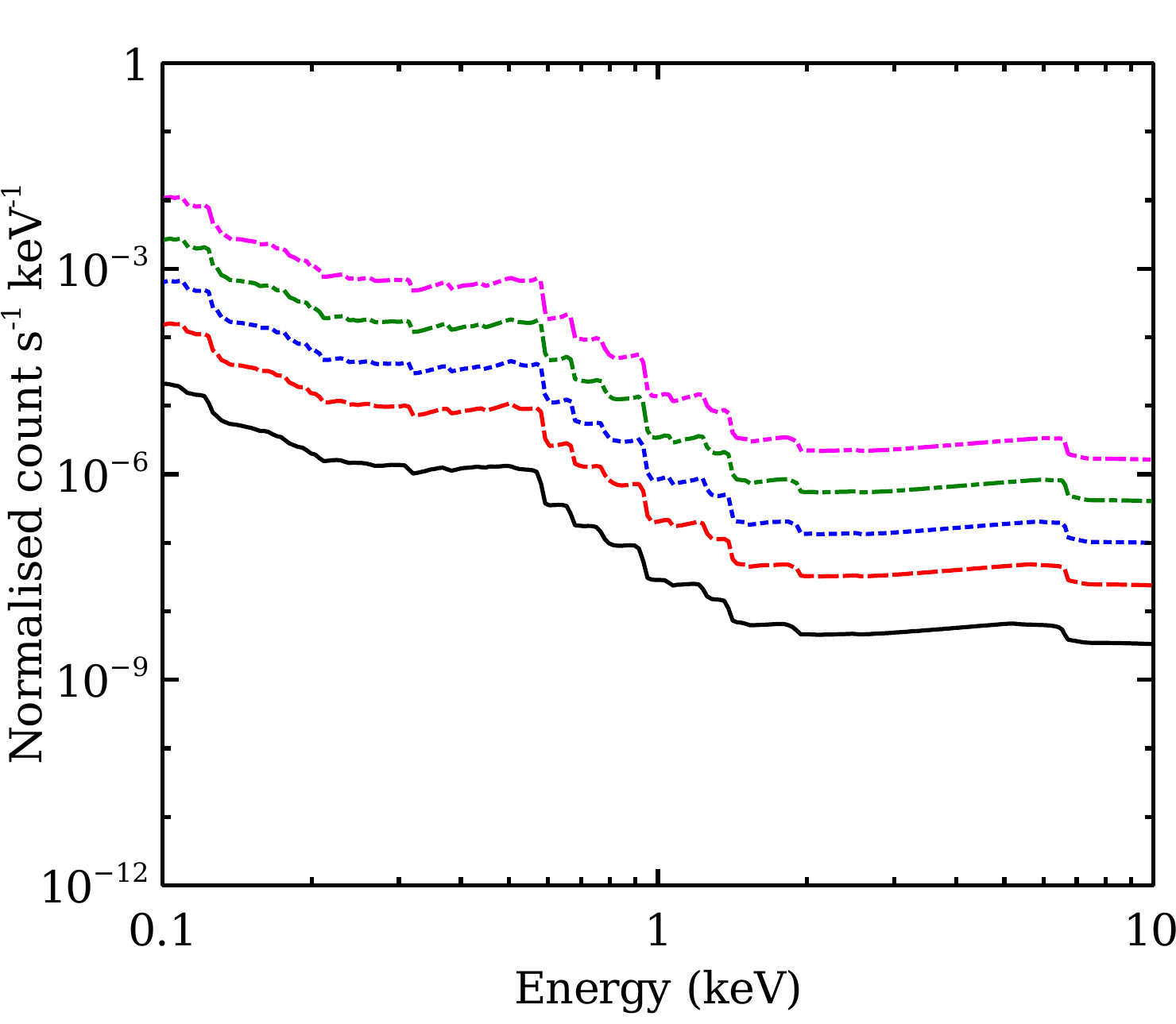}
\caption{Outer radius variation for the low ionized disc ($\xi < 10 \text{ erg cm s}^{-1}$ and $A = 1$) illuminated by an on-axis point source with $h = 5 r_g$ and $\Gamma = 2$. Other parameters are $a = 0.998$, $\theta = 30^\circ$ and $R_S = 1$. \emph{Top panel:} The reverberation lags for five values of the outer radius. \emph{Bottom panel:} the corresponding reflection spectra. While changing the outer radius does not significantly affect the shape of the reflection spectrum, discs with a large outer radius show longer time lags. \label{fig:outer_radius}}
\end{figure}

Although elsewhere in this paper we assume the outer radius of the disc $r_o = 400 r_g$, here we investigate the effects of relaxing this assumption. Time lags and the reflection spectra for variable $r_o$ are presented in Fig.~\ref{fig:outer_radius}. When the disc is extended further out, there are more regions for reflection associated to longer time delays. Therefore the averaged response time increases and hence time lags increase. We can see the maximum lag difference between $r_o = 20$ and $400r_g$ is $\approx 5 t_g$. However, their reflection spectra are almost identical which agree with the theoretical framework in which the prominent fluorescent lines (e.g., the broad Fe K$\alpha$ line) are produced at the inner disc so they are almost independent on $r_o$.

It is clear that while the parameter $r_o$ has less effects on the averaged spectrum, it does affect time lags. There is also the possibility that the response fractions in soft and hard bands increase with $r_{o}$ because of an extended reflecting area. Since the outer regions have  a lower ionization state than the inner regions, the ratio $R_S / R_H$ is larger for the outer part of the disc. If this is the case, the maximum lag differences are larger than those in Fig. 5, which assumes equal $R_{S}=1$ for all cases. However, to limit the number of parameters we restrict our consideration to $r_o = 400 r_g$. We note that $r_o$ could affects time lags, and could be included as a variable parameter in the future.

\section{Fitting Mrk 335} \label{sec:model_fits}

\subsection{Model parameters}

We consider a maximally spinning black hole and assume the accretion disc extends from $r_{ms}$ to $400 r_g$. The cut-off energy of the X-ray continuum is fixed at 300~keV. Our model (so-called {\sc revb} model) is coupled to the {\sc xillver} model that has variable angles of reflection X-rays with respect to the disc's normal between $5^\circ - 85^\circ$. For simplicity only one value of inclination $\approx 50^\circ$ is selected but this could be relaxed in the future. We produce the grid of parameters shown in Table~\ref{tab:model_parameters} which consists of 40,320 grid cells covering possible values that have been found by previous studies. Each grid cell corresponds to a given set of parameters: the source height ($h$), disc inclination angle ($i$), photon index ($\Gamma$), iron abundance ($A_{Fe}$), ionization state at the innermost stable circular orbit ($\xi_{ms}$), the disc density index ($p$) and the soft reflected response fraction ($R_{S}$). A model without the response fraction is very challenging and depends upon many factors (e.g. the efficiency of reflection or the normalisation of the models). Although ray tracing simulation allows us to count the number of hard photons that escape to the observer and that irradiate the disc, the proportionality coefficient of the reflection and the incident flux cannot be determined without further assumption of such as the albedo of the reflecting disc. This makes calculating the exact value of the response much more complicated. For this reason we allow the fraction $R_{S}$ to vary and measure the corresponding $R_{H}$. In other words the $R_{S}$ is one of the model parameters which once selected will determine a unique value of $R_{H}$. Note that our model grids extend over a large, multi-dimensional parameter space which is computationally difficult to interpolate over, and when fitting data will result in a complex $\chi^2$ space that will have many local minima. We therefore have chosen to step through the entire grid of models and allow, for example, the black hole mass and normalisation factors to be free parameters.

\begin{table}
\begin{tabular}{ l l }
\hline
\multirow{2}{*}{Parameters} & \multirow{2}{*}{Value} \\
 & \\ \hline
$h$ ($r_g$) & 2, 3, 4, 5, 6, 7, 8 \\
$i$ ($^\circ$) & 15, 30, 45, 60  \\
$\Gamma$ & 1.8, 2, 2.2, 2.4 \\
$A_{Fe}$ & 0.5, 1, 2 \\
log $\xi_{ms} (\text{ erg cm }s^{-1})$ & 1, 2, 3, 4, 5 \\
$p$  & 0, 1, 2 \\
$R_S$ & 0.3, 0.4, 0.5, 0.6, 0.7, 0.8, 0.9, 1 \\
\hline
\end{tabular}
\caption{Grid of parameters for which the {\sc revb} model has been computed. \label{tab:model_parameters}}
\end{table}

The models that we have computed predict both the time-averaged spectrum and reverberation lags, so they can be simultaneously fit to the data. Since our reverberation lag model does not explicitly compute the positive lags, we employ a power-law ({\sc powerlaw} model) in the form of $\tau(f)=N(cf)^{-s}$ to produce the positive lags where $N \in [0, 10^6]$, $s \in [0,6]$ and $c = 10^4$ is a scale factor of the frequency $f$. Therefore the total timing model is $\textsc{revb} + \textsc{powerlaw}$. The time in gravitational units is transformed to physical units for the variable black hole mass between $10^6 - 10^9 M_\odot$. For the spectral model we add two narrow Gaussain lines at $\approx$ 6.4 and 7 keV responsible for the refection from very distant material as discussed in \citet{On07} and \citet{La08}, and necessary additional components when modelling the broad iron line. These narrow features do not affect the lags as they do not vary on the timescales of the inner disc reflection. All spectral components are modified by galactic absorption with $N_H = 3.99 \times 10^{20} \text{ cm}^{-2}$. The total spectral model is produced via $\textsc{tbabs} \otimes (\textsc{revb} + \textsc{gauss1} + \textsc{gauss2})$.

\subsection{Results}

We begin by fitting the 2--10~keV spectrum and the 2.5--4 vs. 4--6.5 keV time lags in high flux state of Mrk 335. The fitting procedure is performed in {\sc ISIS} \citep{Ho00} using the {\sc subplex} minimisation method. In {\sc ISIS} we load the time-averaged \emph{XMM-Newton} EPIC-pn spectrum, background and responses as the first data set. We then define a second data set which represents the time lag vs. frequency, rather than the standard counts vs. wavelength. For a given set of parameters our reverberation model produces the time-averaged spectrum which is fit to the first data set and the time lag spectrum that is, simultaneously, fit to the second data set. The reported $\chi^2$ values of fit statistics are for both data sets combined. We have been able to achieve excellent fits without having to change the relative weighting of the spectroscopic and timing data sets, although this can be done if required.

\begin{figure}
\centering
 \vspace{-0.7cm}
\includegraphics*[width=85mm]{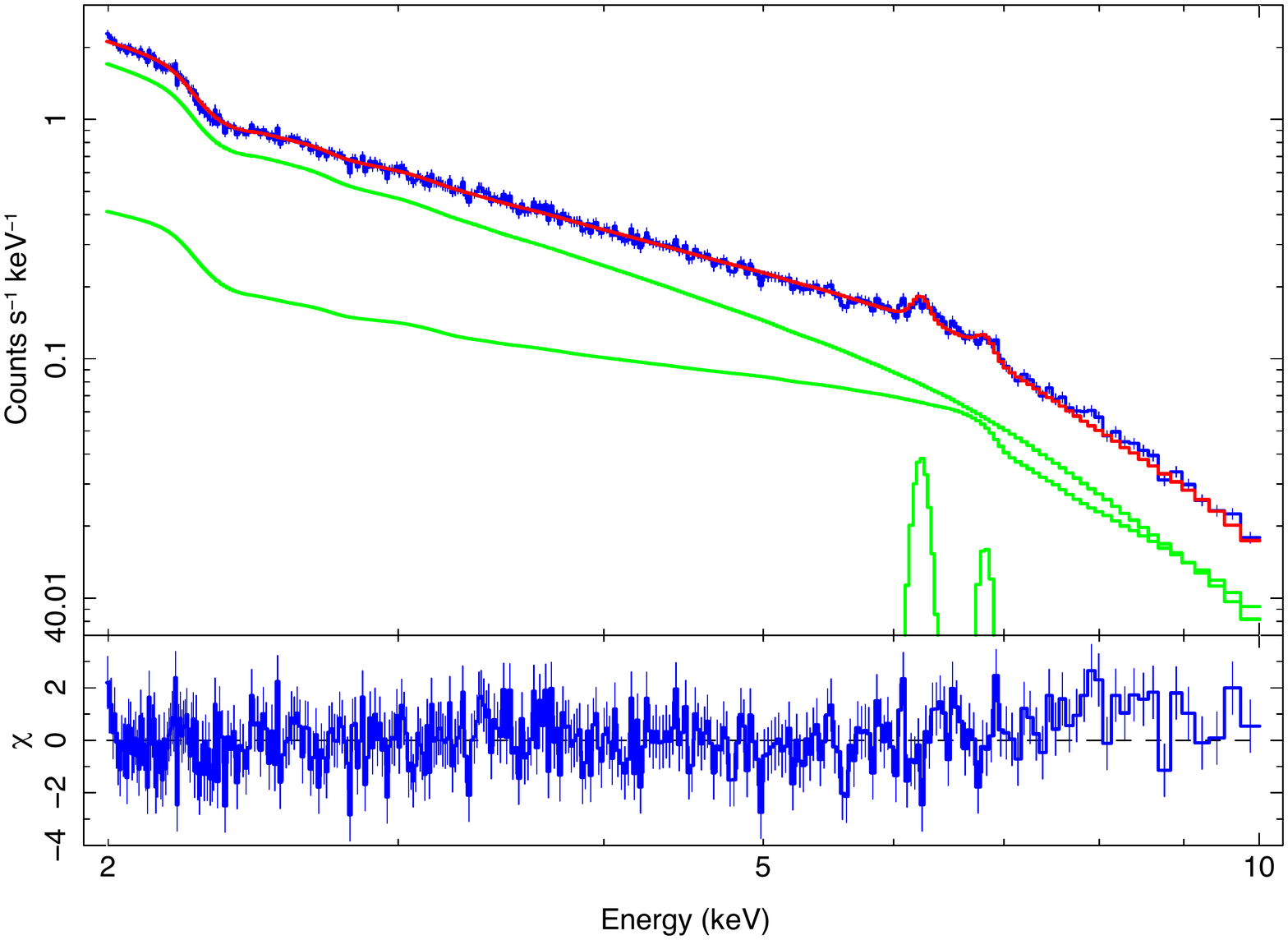}
\vspace{-0.3cm}
\includegraphics*[trim = 0mm 0mm 0mm 25mm, clip, width=85mm]{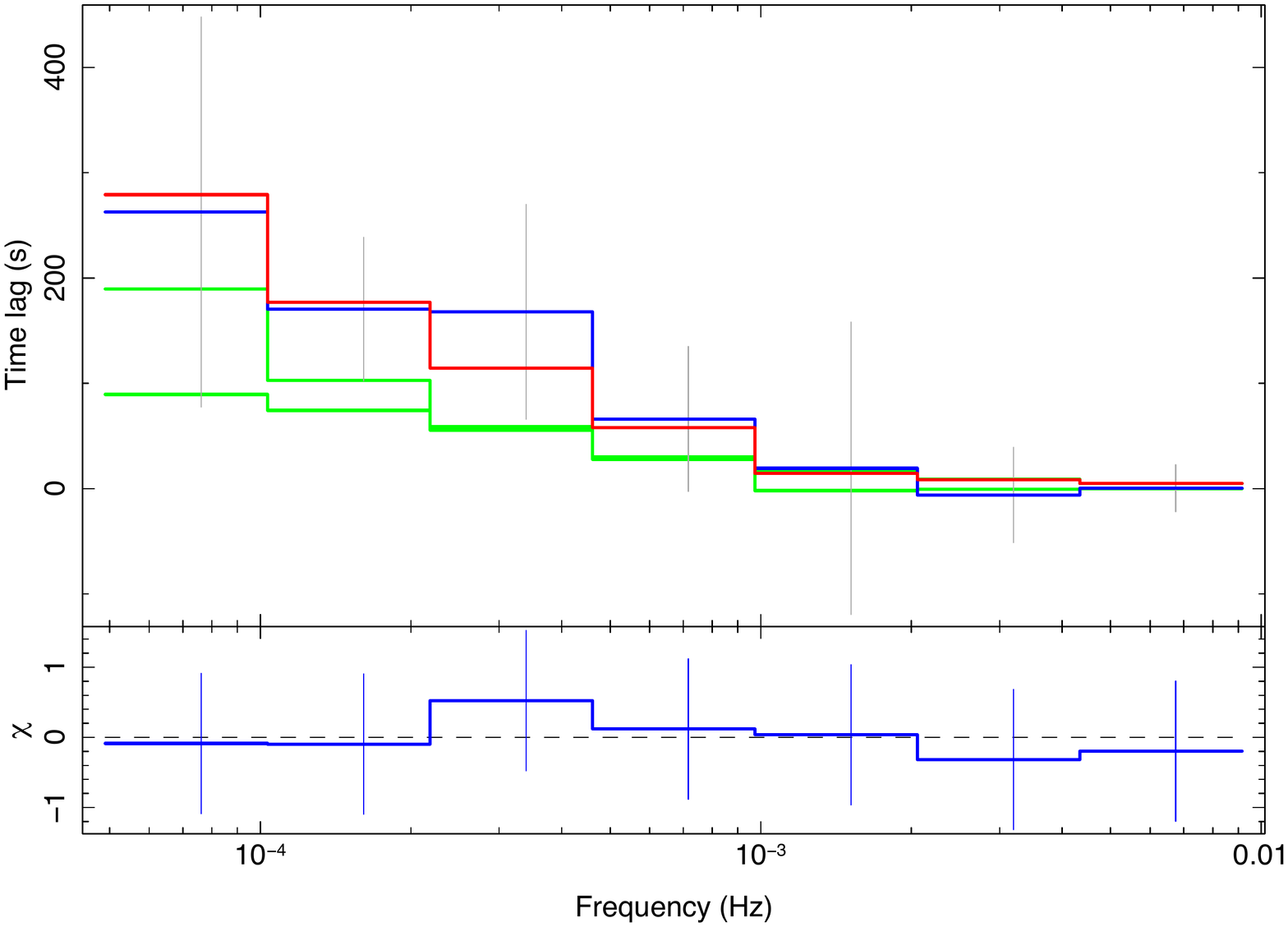}
\caption{Data and residuals (blue points) from simultaneously fitting the $\textsc{revb} \otimes \textsc{xillver}$ to the high flux state spectra (\emph{top panel}) and time lags between 2.5--4 and 4--6.5 keV bands (\emph{bottom panel}). The overall model is shown in red, which is the sum of individual model components (shown in green). The individual spectral components are the continuum, blurred reflection, and two iron lines (from top to bottom, respectively) and the individual lag components are the power-law and reverberation model (from top to bottom, respectively). Note that there is clearly a degeneracy between the the power-law, which may be attributed to the propagation of fluctuations in the disc, and the reverberation signature. \label{fig:hb-fits}}
\end{figure}

The best fit spectral and timing profiles are shown in Fig.~\ref{fig:hb-fits}. The best-fitting parameters are listed in Table~\ref{tab:best_fit}. The model provides excellent fits to the data with $\chi^2 / \text{d.o.f.} = 1.026$. The X-ray source is at $2r_g$ on the symmetry axis. The time lags are well-fitted for the black hole mass $M \approx 1.3 \times 10^7 M_\odot$. The accretion disc has an inner ionization parameter $\xi_{ms} = 10^{3} \text{ erg cm s}^{-1}$, $A_{Fe} = 0.5$ and $p = 0$. The inclination angle is $45^\circ$ and the soft reflected response fraction is 0.3. We find the photon index of the continuum X-rays is 2.4 steeper than those of normal AGN. However, the steep photon index has been found before in Mrk 335 by such as \cite{Go02} and \cite{Wi15}. The fits are poorer without two narrow Gaussians at $\approx 6.4$~keV (Fe K$\alpha$) and $7$~keV (Fe XXVI). While the first line can be interpreted as the result of reflection from distant, cold material such as a molecular torus, the second line may originate in hot, optically thin material such as the hot gas filling that torus. Note that the analysis of \citet{On07} required these two additional narrow line components in addition to the relativistically broadened reflection component. These narrow features from distant reflection do not vary on the timescales of the inner disc reflection so including them in the spectral model do not have any effect on the lags. Therefore the model can provide a self-consistent fit to both spectrum and time lags within the same energy band.

\begin{table}
\begin{tabular}{ l l l }
\hline
 \multirow{2}{*}{Component} & \multirow{2}{*}{Parameter} & \multirow{2}{*}{Value}\\
 & \\ \hline
{\sc tbabs}$^{s}$ & $N_H (\times 10^{20} \text{ cm}^{-2})$ & $4.0^f$ \\
{\sc revb}$^{s+t}$	& $h (r_g)$ & 2 \\
		& $i$ ($^{\circ}$) & 45  \\
		& $\Gamma$  & 2.4 \\
		& $A_{Fe}$  & 0.5 \\
		& log $\xi_{ms} (\text{ erg cm s}^{-1})$ &  3\\
		& $p$ &  0\\
		& $R_S$ & 0.3 \\
		& Norm1 & $7.6 \times 10^5$ \\
		& log $M (M_\odot)$ &  7.1\\
{\sc powerlaw}$^t$ & $s$ & 0.82 \\
		& Norm2 &  151.8\\
{\sc gauss1}$^s$ & Area ($\times 10^{-6}$)	& $9.81^f$  \\
        & Centre (keV) & $6.4^f$ \\
{\sc gauss2}$^s$ & Area ($\times 10^{-6}$)	& $4.80^f$  \\
        & Centre (keV) & $7.0^f$ \\
$\chi^{2} / \text{d.o.f.}$ & & 1.026 \\
\hline
\end{tabular}
\caption{Model parameters for fits to the 2--4 keV spectrum and time lags between 2.5--4 and 4--6.5 keV bands of Mrk 335 during the high flux state. The model components, their parameters and the parameter values are listed in Columns 1, 2 and 3, respectively. The superscript $s$ refers to spectral parameters while $t$ refers to timing parameters. The superscript $f$ identifies the parameters which are fixed. Line energies refer to the rest-frame. \label{tab:best_fit}}
\end{table}

Having successfully fitted a physical reverberation model to the 2--10 keV spectrum and time lags of 2.5--4 vs. 4--6.5 keV bands, we now extend to fit the spectrum down to 0.3 keV with the Fe-L lags of 0.3--0.8 keV vs. 1--4 keV bands. To deal with the soft excess at $\approx$ 0.3--2 keV which is often seen in AGN, we replace two Gaussians with more physical models, unblurred $\textsc{xillver}$ and $\textsc{vmekal}$, responsible for distant reflection from cold medium and ionized diffused-gas, respectively. The warm absorber as reported by some previous studies \citep[e.g.][]{Lo13} is modelled by $\textsc{xstar}$ \citep{Kallman2001}. Finally, a blackbody component with a maximum temperature fixed at 0.05 keV is added and is corresponding to the disc radiation which is constant over time. Therefore, all components we include so far should have no (or very little) effect on the reverberation lags.

\begin{figure}
\centering
\vspace{-0.7cm}
\includegraphics*[width=85mm]{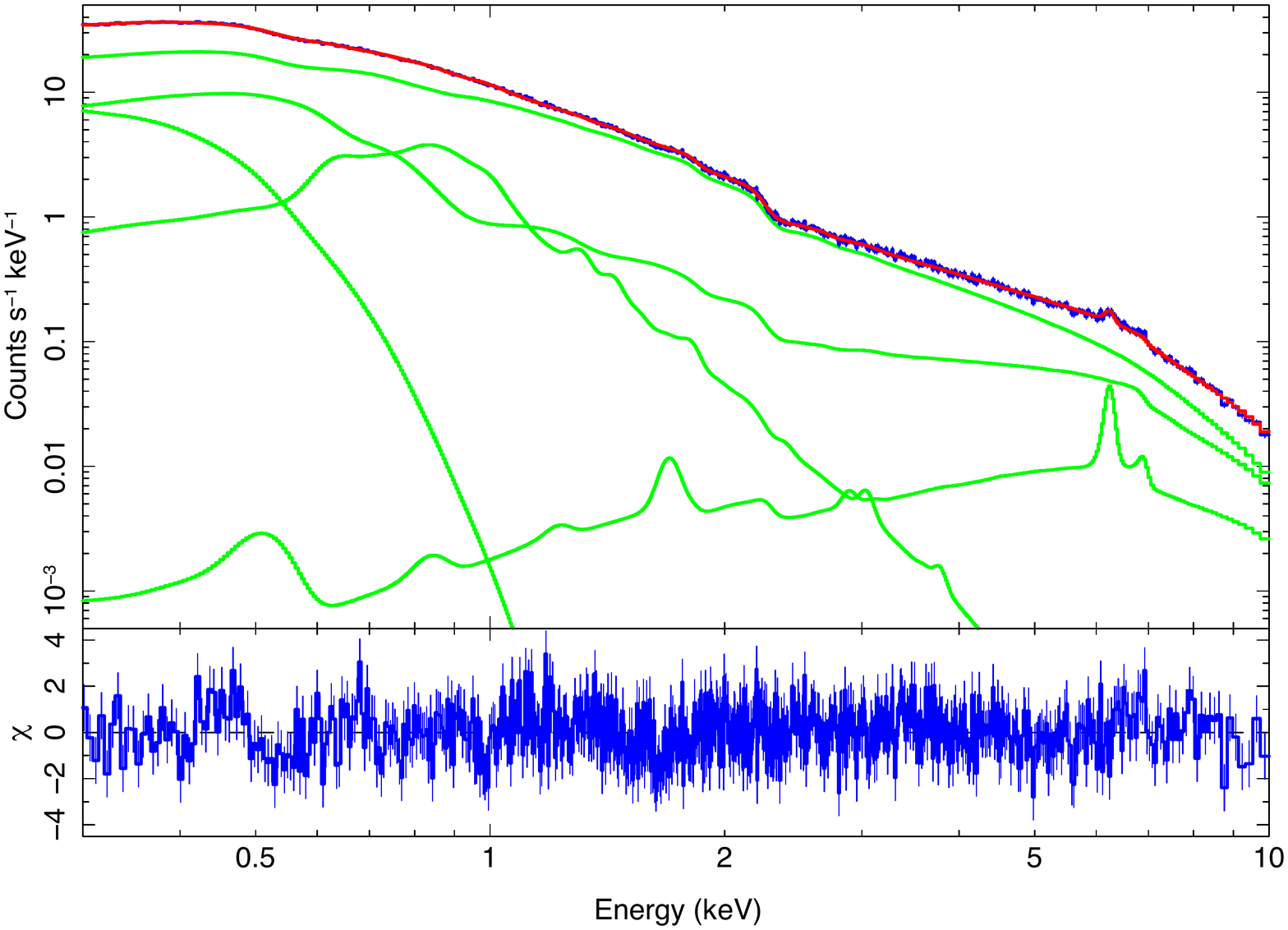}
\vspace{-0.3cm}
\includegraphics*[trim = 0mm 0mm 0mm 25mm, clip, width=85mm]{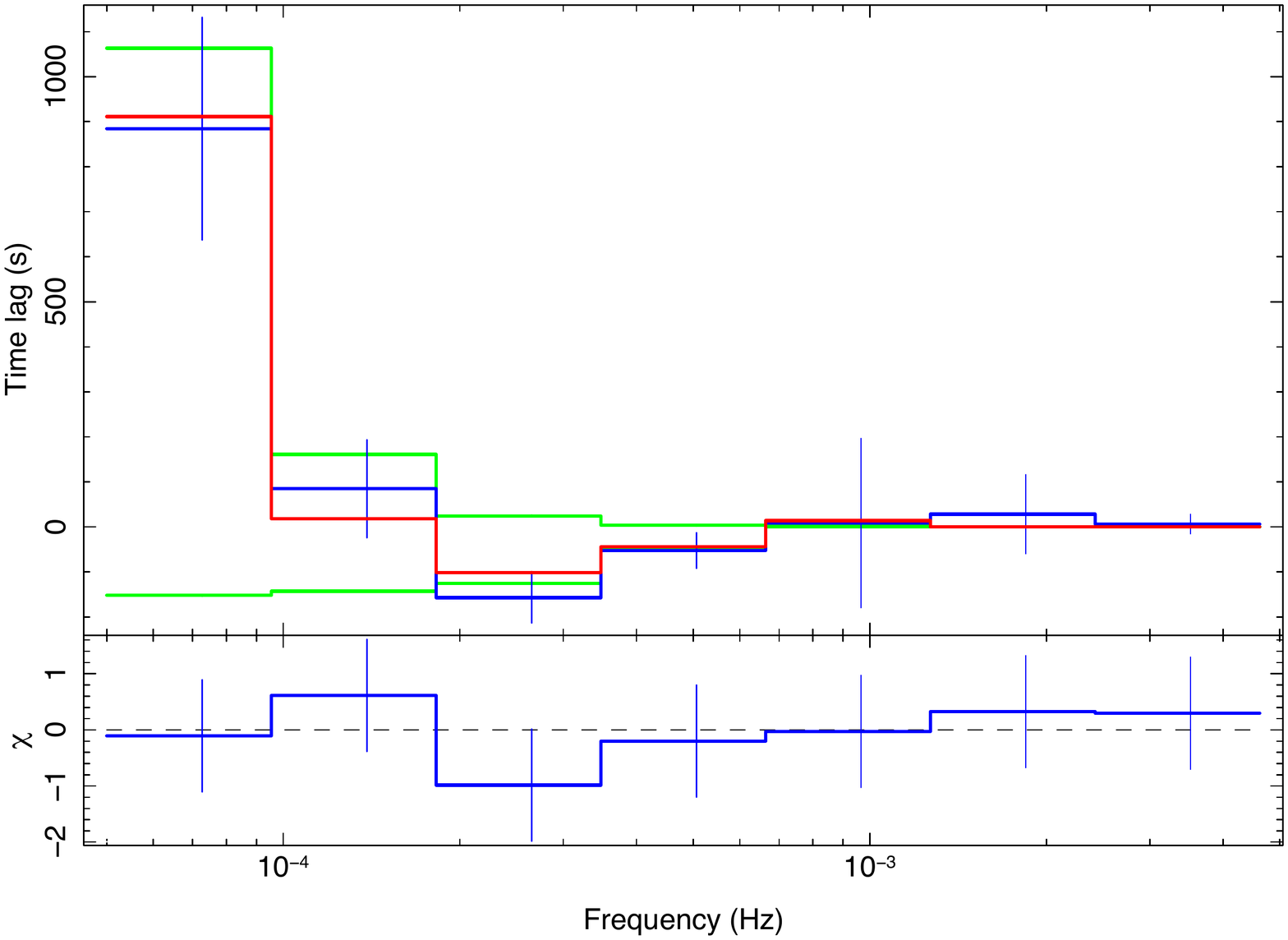}
\caption{Data and residuals from simultaneously fitting the $\textsc{revb} \otimes \textsc{xillver}$ to the 0.3--10 keV spectra (\emph{top panel}) and the time lags between 0.3--0.8 vs. 1--4 keV band representing the Fe L lags (\emph{bottom panel}). The individual model components are shown in green. In the top panel, the spectral components, folded through the instrument response, are the continuum, blurred reflection due to reverberation, and {\tt xillver} neutral reflection (from top to bottom, respectively, at high energies), a thermal component that exceeds the blurred reflection component around $1 \text{ keV}$, and a blackbody component that only contributes significant flux $\lesssim 0.5 \text{ keV}$. In the bottom panel, the lag components represent power-law and reverberation (from top to bottom, respectively). In contrast to the case shown in Fig.~\ref{fig:hb-fits}, the power-law and reverberation lag components have opposite signs so are easier to separate. \label{fig:sb-fits}}
\end{figure}

The model fitting to 0.3--10 keV spectrum and Fe L lags still provides good fits to the data with $\chi^2 / \text{d.o.f.} = 1.15$ (see Fig.~\ref{fig:sb-fits}). Previous studies on the RGS spectrum \citep[e.g.,][]{Gr08, Longinotti2008} found the atomic emission lines from the  ionized plasma play an important role in the soft bands. In the $\textsc{vmekal}$ model, we allow the abundances of O, Si, S and Fe atoms to vary between 0.2--5, and tied the abundances of other atoms together. The important parameters are listed in Table~\ref{tab:soft_band_fit}. Most of the {\sc revb} parameters are comparable to what previously obtained by constraining the lags in harder energy band except $\xi_{ms} = 100 \text{ erg cm s}^{-1}$, $p=2$ and $R_{S}=0.4$. The difference of $R_{S}$ can be understood as the selected soft bands are different. The reverberation lags in this case are negative since the soft band is more dominated by the reflection flux comparing to the hard band. The lags scale with the mass $M \approx 2 \times 10^7 M_\odot$. The {\sc powerlaw} components seen in both cases are quite different. This is because the continuum lags are dependent on the selected energy band as same as the reverberation lags.

\begin{table}
\begin{tabular}{ l l l }
\hline
\multirow{2}{*}{Component} & \multirow{2}{*}{Parameter} & \multirow{2}{*}{Value}\\
& \\ \hline
{\sc tbabs}$^{s}$ & $N_H (\times 10^{20} \text{ cm}^{-2})$ & $4.0^f$ \\
{\sc xstar}$^{s}$	& Norm       & $1.7 \times 10^{21}$ \\
& log $\xi_{ms} (\text{ erg cm s}^{-1})$ &  1.85\\
{\sc revb}$^{s+t}$	& $h (r_g)$ & 2 \\
		& $i$ ($^{\circ}$) & 45  \\
		& $\Gamma$  & 2.4 \\
		& $A_{Fe}$  & 0.5 \\
		& log $\xi_{ms} (\text{ erg cm s}^{-1})$ &  2\\
		& $p$ &  2\\
		& $R_S$ & 0.4 \\
		& Norm1 & $5.6 \times 10^5$ \\
		& log $M (M_\odot)$ &  7.3\\
{\sc powerlaw}$^t$ & $s$ & 2.92 \\
		& Norm2 &  419.6\\

		{\sc blackbody}$^s$ & $T$ (keV) & 0.05 \\
		& Norm &  $4.0\times 10^{-4}$\\

unblurred {\sc xillver}$^s$   	& log $\xi_{ms} (\text{ erg cm s}^{-1})$ &  $0^f$\\
	        	& $A_{Fe}$       & 0.7 \\
{\sc vmekal}$^s$    & $T$ (keV)	        & 0.51 \\
                    & $A_{O}$           & 2.8 \\
	        	    & $A_{Si}$          & 0.2 \\
	        	    & $A_{S}$           & 0.2 \\
	        	    & $A_{Fe}$          & 0.9 \\
	                & $A_\text{other}$  & 1.8 \\

$\chi^{2} / \text{d.o.f.}$ & & 1.15 \\
\hline
\end{tabular}
\caption{Model parameters for fits to the 0.3--10 keV spectrum and Fe L lags between 0.3--0.8 vs. 1--4 keV bands of Mrk 335 during the high flux state. The model components, their parameters and the parameter values are listed in Columns 1, 2 and 3, respectively. The superscript $s$, $t$ and $f$ refers to spectral parameters, timing parameters and fixed parameters, respectively. \label{tab:soft_band_fit}}
\end{table}

\subsection{Discussion}

Simultaneously fitting the 2--10 keV spectrum and time lags between 2.5--4 keV and 4--6.5 keV is straightforward for two main reasons. Firstly, the lags are in the same energy band as the time-integrated energy spectrum. Secondly, two additional narrow features $\approx 6.4$~keV and $7$~keV from the distant reflection do not affect the lags because they do not vary on the reverberation timescales. However, shifting the lags to fit the Fe L band is much more difficult due to the soft excess at $\approx$ 0.3--2 keV and the choices of appropriate additional components/models. One important reason why the $\textsc{revb}$ model alone does not fit the soft band well is that while the iron abundance was allowed to vary the abundances of other elements were set to the Solar value. Allowing variable abundances of other atoms in the $\textsc{revb}$ model should significantly improve the fits in the 0.3--2 keV band but this will make changes in the soft response fraction and reverberation lags. Unfortunately these changes are beyond the scope of our present paper, although the change to the soft band spectrum could be significant (cf. the discussion in the Appendix on differences between reflection model with the \emph{same} abundances). Instead we model the distant reflection using (unblurred) $\textsc{xillver}$, and the soft excess with a $\textsc{vmekal}$ plus $\textsc{blackbody}$ component at the maximum temperature 0.05 keV. These additional components do not in principle significantly affect time lags. Future high resolution spectroscopy with \emph{Astro-H} and \emph{Athena} will be particularly helpful in disentangling the spectral complexity of the soft X-ray band.

We find that one warm absorber is needed for the good fit below 2 keV. The unblurred $\textsc{xillver}$ contributes mainly to the narrow line at $\approx 6.4$ keV but very little to the rest of spectrum, so using a Gaussian is still a good approximation. The plasma temperature obtained with the $\textsc{vmekal}$ model is low (0.51~keV) so it does not provide a perfect fits to the $\approx 7$ keV weak Fe XXVI line. However, it does remarkably improve the fits below 2 keV. The requirement of $\textsc{xillver}$ (or $\approx 6.4$ keV Gaussian line) and $\textsc{vmekal}$ suits the framework of the reflection from distant cold-molecular torus and the ionized gas filling the torus \citep{On07}. This is a plausible model to fit the 0.3--10 keV spectrum without invoking the components that might significantly affect time lags.

The soft excess can also be modelled by either a blurred reflection or a $\textsc{blackbody}$ component with a temperature $\approx$ 0.1--0.2 keV \citep{Gi04,Cr06}. Although this temperature is argued to be too hot for the thermal emission from a standard disc, it is possible that the inner disc under strong light bending effects is heated by the returning radiation, causing the disc to emit the soft X-rays. \cite{Gr08} successfully modelled the high flux state of Mrk 335 using two power-laws, either two or one corresponding blurred inner-disc reflection, plasma emission by $\textsc{mekal}$ model and some further emission lines. Recently \cite{Wi15} fitted the same spectrum profile but between 0.5--10 keV using the blurred and unblurred components modified by the warm absorbers. Our  $\textsc{blackbody}$ component is negligible if we fit the 0.5--10 keV spectrum and time lags ($\chi^{2}/d.o.f. \approx 1.1$) because it contributes to the flux $<5$ keV. Using $\textsc{mekal}$ rather than $\textsc{vmekal}$ gives acceptable but poorer fits ($\chi^{2}/d.o.f. \approx 1.3$). Keeping in mind the complexity of the soft excess (e.g. its uncertain origin and variation-timescales), our current model is more self-consistent if the spectrum and the lags are fitted above the 2 keV band.

Moreover, we find that the model still provides a good description of the data ($\chi^{2} / \text{d.o.f.} <1.2$) for all disc density indices ($p=$ 0, 1 and 2), but the source height and the black hole mass can be different. Of the values we investigated, the source is at $\approx 2-3 r_{g}$ on the symmetry axis and the acceptable black hole mass is $\approx 1.3-2.5 \times 10^7 M_\odot$ depending on a given density profile of the disc. Our constrained mass agrees with the estimated mass that follows the mass-scaling laws \citep[e.g.][]{De13}. It is in good agreement with the masses reported by \cite{Grier2012} and \cite{Em14} derived from optical and X-ray reverberation, respectively. The source height is slightly lower than \cite{Em14} where $h=3.1r_{g}$ was found by fitting only the Fe-L lags and neglecting the contamination in the hard band. Such low height means the corona is very compact even in the high flux state of this source. It would be interesting to see how the model parameters change when the source is in a low flux state.

Simultaneously fitting spectral and timing data may help break degeneracies between the continuum source height and black hole mass. The source height determines the illumination pattern and, given the disc density profile, the corresponding reflection spectra. The black hole mass sets the light-crossing timescale of a gravitational radius, and the source height gives the associated reverberation lags. Only by combining spectroscopy and timing products are both the black hole mass and source height constrained. However, different disc density profiles may result in different estimates of the source height and black hole mass. This suggests that a better understanding of the radial density distribution in the disc is needed which in turn allows a more accurate estimation of the ionization gradient corresponding to a particular illuminating X-ray source. Otherwise there will be some degeneracy between the source height and central mass produced by different density profiles that still provide excellent fits to the data. Fitting energy-dependent time lags (e.g., the full lag-energy spectrum) might break these degeneracies, which will be investigated in future papers.

Finally, it is important to note that even considering reverberation between harder X-ray bands is satisfactory without taking into account the uncertainty of the ``soft excess'' below 2 keV (or, crucially, how it varies), the sign of the time lag caused by putative fluctuations propagating through the disc and reverberation due to light travel time are the same (see the green lines in the bottom panel of Fig.~\ref{fig:hb-fits}). This possible degeneracy will be addressed in a future paper by self-consistently modelling the propagating fluctuations. This also points towards joint spectral and timing modelling being more straightforward, from a theoretical perspective, in the Fe K band which is observationally more challenging but ideally suited to the \emph{Athena} X-ray observatory \citep[e.g.,][]{Dovciak2013}.

\section{Conclusions} \label{sec:conclusions}

We present a self-consistent model for simultaneously fitting time-averaged and lag-frequency spectra in the high flux state of Mrk 335. To reduce the number of free parameters in our initial model, we restrict our consideration to the lamp-post geometry with a maximally rotating black hole with a standard disc extending between $r_{ms}$ and $400r_g$. The model including full-dilution effects is successfully fitted to the combined 2--10 keV spectrum and the time lags of 2.5--4 vs. 4--6.5 keV bands. We find the inclination angle is $45^{\circ}$, the iron abundance is 0.5 and the photon index of direct continuum is 2.4. The X-ray source height is very small, $\approx 2r_{g}$, illuminating a constant density disc ($p=0$). The ionization parameter is $10^{3}$ erg cm s$^{-1}$ at the innermost part of the disc and decreases further out. The black hole mass is $\approx 1.3 \times 10^{7}M_{\odot}$. Our results also show that, along with the best value of $p$, the model still provides excellent fits but the constrained source height and black hole mass are different. This implies that the well-define density profile is needed in order to robustly determine the source height and the central mass.
Having successfully fitted a physical reverberation model to the spectrum and time lags in the hard band, we discuss the difficulties of fitting the soft band due to the soft excess since its origin is unclear. Nevertheless, we have presented a plausible model that successfully fit this band by further adding only the components which have no (or less) effects on reverberation lags. These components, unblurred $\textsc{xillver}$ and $\textsc{vmekal}$, confirm an existence of both cold and ionised gas at very distant. In the Appendix, the mismatch of soft-excess between {\sc xillver} and {\sc reflionx} models is also reported which makes the interpretation of the physics at 0.3--2~keV much more complicated. This is however worth investigating in the future.

Furthermore the positive lags which we are modelling phenomenologically by a power-law could be replaced by the hard lags from a physical model. The choice of mechanisms that drives positive lags (e.g. propagating fluctuations on the disc) is also important. Modelling the propagation of accretion rate fluctuations and an associated spatially extended corona geometry is planned for a future paper, but is significantly beyond the scope of this paper. This is particularly important for constraining the otherwise fairly arbitrary power law that can be applied to the frequency dependent time lags.

The model could also be improved by allowing the outer radius to be a free parameter as it affects reverberation lags (see Fig.~\ref{fig:outer_radius}). Additionally, we plan to include the energy-dependent time lags in our fitting procedure. All are possible and straightforward but, undoubtedly, make the model more computationally intensive. Fitting combined spectral and timing data is important as the parameters are better constrained than fitting each one alone. It also has the potential to constrain or measure the spectrum of reflection models in the soft X-ray band.

Finally, there are two spectral components that our model does not capture -- reflection from distant neutral material, and thermal emission from some hot gas whose origin is less clear. Both additional components are needed in ``traditional'' models of the broad iron line in time-averaged spectra. Future observations with micro-calorimeters will really help constrain the location and dynamics of the gas associated with these narrow lines. The features from distant reflection, however, do not vary on the timescales of the inner disc reflection which make the reverberation technique very powerful.

We have demonstrated the feasibility of successfully fitting cross spectra to constrain the corona and disc geometry. These initial fits reveal some degeneracies between parameters that can be investigated by looking at different flux states and by analysing a sample of objects. A more sophisticated, and likely spatially extended, model must be developed to self-consistently explain the positive lags at low frequency. For example, at the moment it is far from clear how the propagating fluctuations connect to the corona. The ``soft excess'' continues to warrant further investigation.

\section*{Acknowledgments}

This work was carried out using the computational facilities of the Advanced Computing Research Centre, University of Bristol\footnote{\url{http://www.bris.ac.uk/acrc/}}. PC thanks the University of Bristol for a Postgraduate Research Scholarship. We thank the referee, Erin Kara, for very useful comments which have improved the paper.

\bibliographystyle{mn2e}
\bibliography{mrk335}

\section*{Appendix}
\renewcommand\thefigure{A\arabic{figure}}
\setcounter{figure}{0}

\begin{figure}
\centering
 \vspace{-0.7cm}
\includegraphics*[width=85mm]{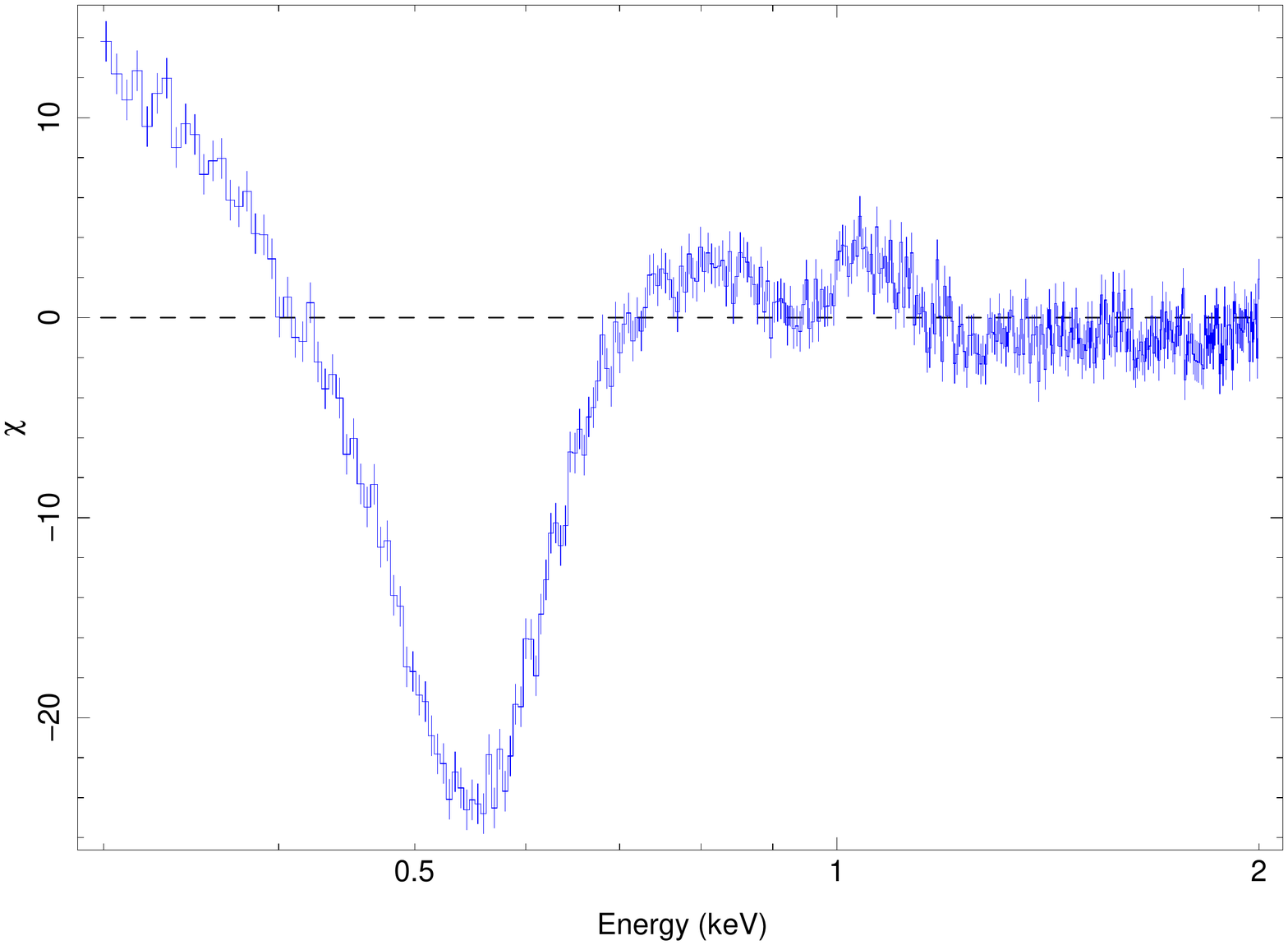}
\vspace{-0.2cm}
\includegraphics*[trim = 0mm 0mm 0mm 25mm, clip, width=85mm]{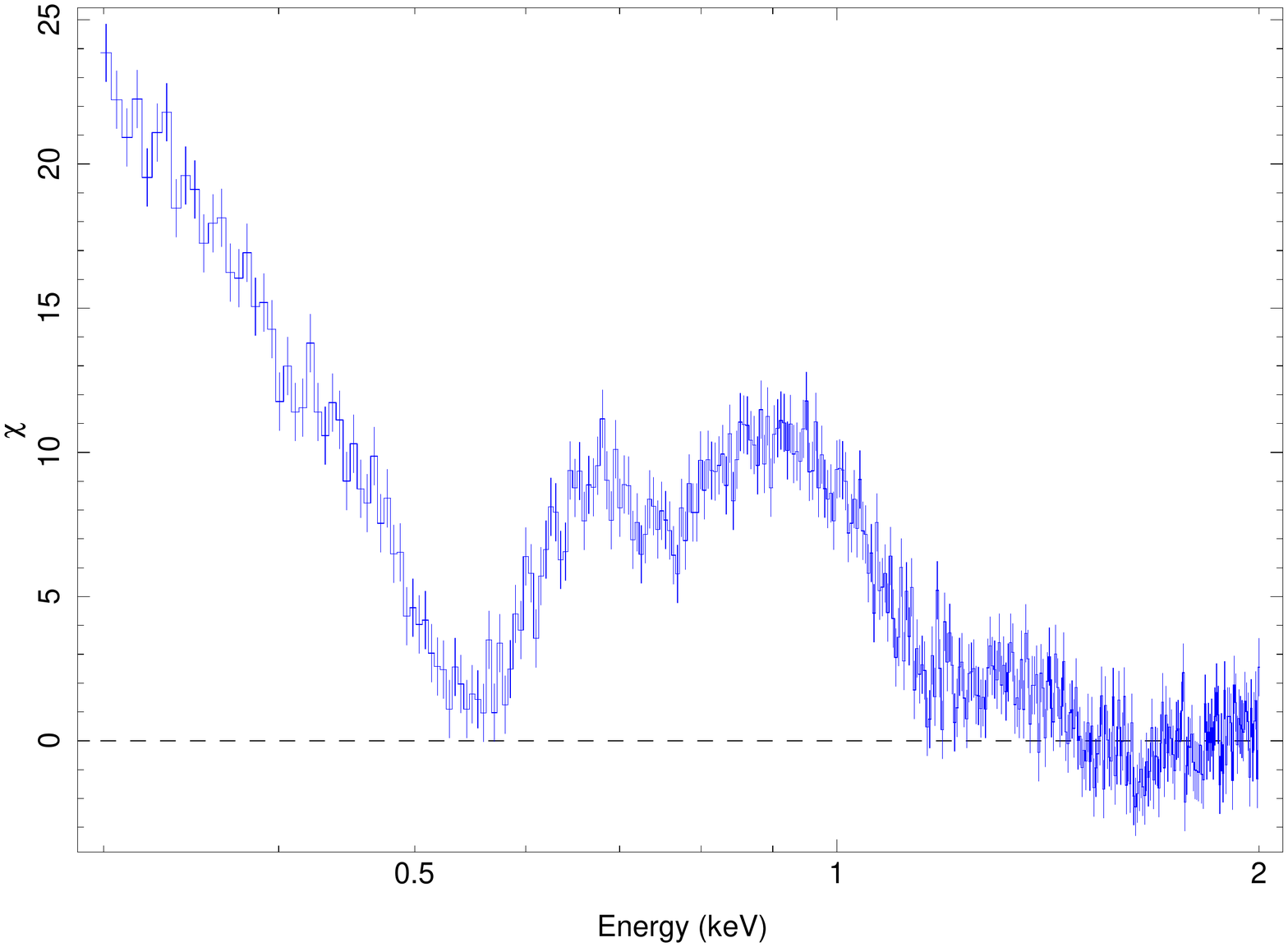}
\caption{Residuals in the 0.3--2.0 keV band for the $\textsc{revb} \otimes \textsc{reflionx}$ (\emph{top panel}) and $\textsc{revb} \otimes \textsc{xillver}$ (\emph{bottom panel}). Each model has been fitted from 2--10 keV adding only two narrow Gaussians at $\approx 6.4$~keV (Fe K$\alpha$) and $7$~keV (Fe XXVI) and then extrapolated to lower energies.  \label{fig:soft_excess_compare}}
\end{figure}

For comparison, and to assess the systematic differences between different reflection models, we couple the {\sc revb} model to the {\sc reflionx} model using similar grid cells as shown in Table~\ref{tab:model_parameters}. The $\textsc{revb} \otimes \textsc{reflionx}$ provides excellent fits to the 2--10 keV spectrum and the hard band lags with $\chi^2 / \text{d.o.f.} = 0.948$. However, the soft excess (Fig.~\ref{fig:soft_excess_compare}) is significantly different to that of the $\textsc{revb} \otimes \textsc{xillver}$. The parameters which have successfully fitted the soft excess for $\textsc{xillver}$ model (Table~\ref{tab:soft_band_fit}) then should not fit for the $\textsc{reflionx}$ model. The mismatch of soft-excess can be understood in term of differences in their atomic data if the {\sc xillver} model is more absorbed at low energies \citep{Gar13}. These systematic differences are certainly larger than our statistical error bars and can affect the interpretation of spectra and time lags.

\begin{figure}
\centering
\includegraphics*[width=65mm]{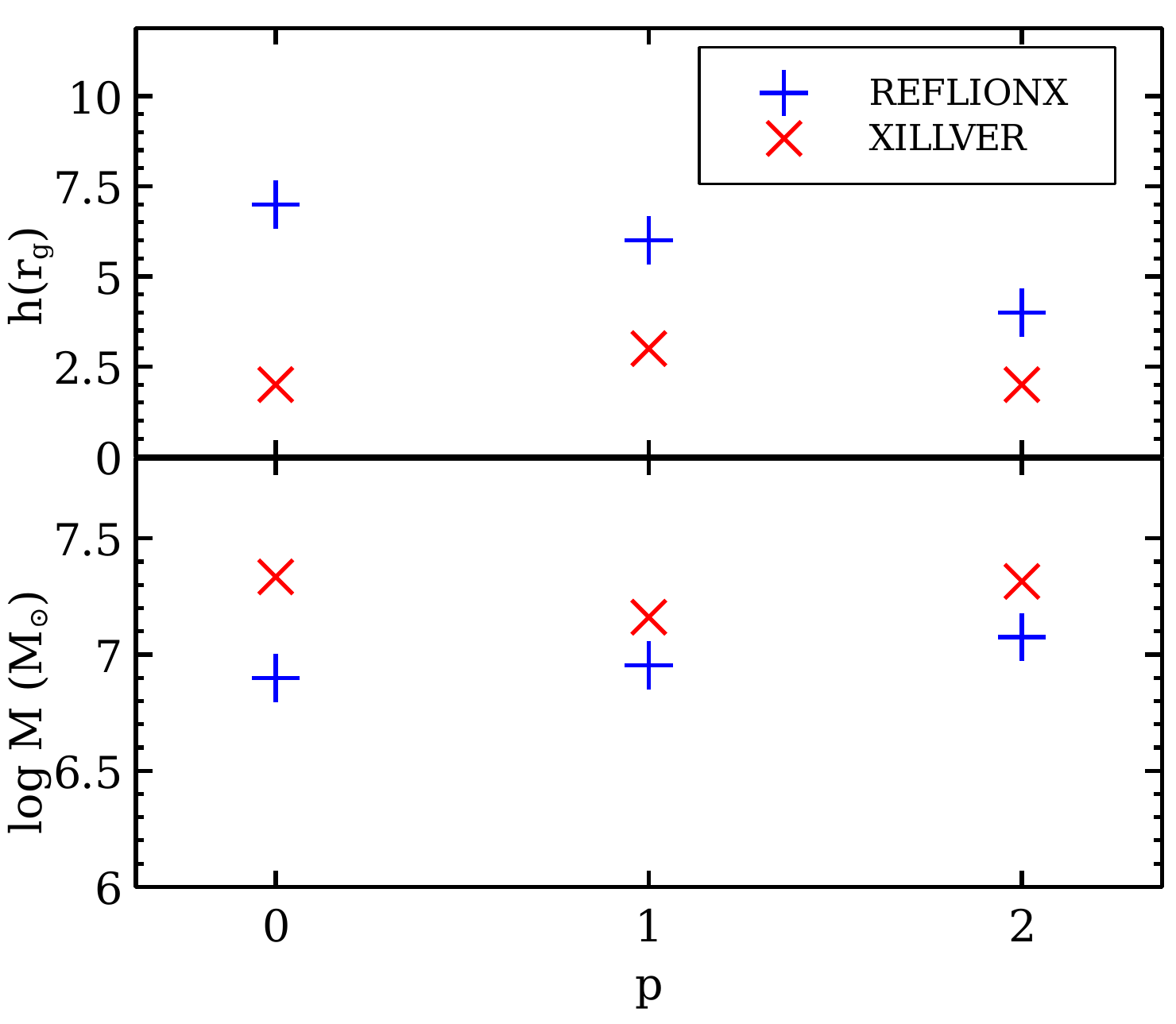}
\vspace{-0.3cm}
\caption{Best fit values of source height (\emph{top panel}) and black hole mass (\emph{bottom panel}) for variable disc-density index, $p$ ($n \propto r^{-p}$). \label{fig:h_and_mass}}
\end{figure}

Keeping in mind the uncertainty of physics below 2 keV, we only compare both models by considering the 2--10 keV spectrum and the hard band lags. It is interesting to note that the best-fit values of the source height and the black hole mass significantly depend not only on the disc density profiles, but also on the choices of reflection models (see Fig.~\ref{fig:h_and_mass}). Within the grid cells we investigate, a constant density disc ($p = 0$) had the largest difference in model parameters. The black hole mass is $\approx 7.9 \times 10^6 M_\odot$ for the {\sc reflionx} model and $\approx 2.0 \times 10^7 M_\odot$ for the {\sc xillver} model. The source's height using {\sc reflionx} model is $7r_g$ which is much larger than in \cite{Em14}. On the other hand, the {\sc xillver} model places the source much lower at $2r_g$. Changing the disc density profile results in a different disc ionization profile, which in turn produces very different reflection spectra and time lags for the two reflection models. The disagreement between the two models, especially when $p = 0$, confirms that in some cases the source height and the central mass are model-dependent. However, only the case of $h=2r_{g}$, as shown in Section~\ref{sec:model_fits}, can provide the best simultaneously fits the time-averaged spectrum and the lags in both soft and hard bands.
\label{lastpage}
\end{document}